\documentclass[12pt]{article}

\author{Yu.~M.~Zinoviev
       \thanks{E-mail address: Yurii.Zinoviev@ihep.ru} \\
        {\it Institute for High Energy Physics} \\
        {\it Protvino, Moscow Region, 142280, Russia}}
\title{On spin 2 electromagnetic interactions}

\date{}

\begin{document}

\maketitle

\begin{abstract}
In this paper we (re)consider the problem of electromagnetic
interactions for massless spin 2 particles and show that in $(A)dS$
spaces with non-zero cosmological constant it is indeed possible (at
least in linear approximation) to switch on minimal electromagnetic
interactions supplemented by third derivative non-minimal ones which
are necessary to restore gauge invariance.
\end{abstract}

\thispagestyle{empty}
\newpage
\setcounter{page}{1}

\section*{Introduction}

It was known for a long time that it is not possible
to construct standard gravitational interaction for massless high spin
$s \ge 5/2$ particles in flat Minkowski space \cite{AD79,WF80,BBD85}
(see also recent discussion in \cite{Por08}). At the same time, it has
been shown \cite{FV87,FV87a} that this task indeed has a solution in
$(A)dS$ space with non-zero cosmological term. The reason is that
gauge invariance, that turn out to be broken when one replace ordinary
partial derivatives by the gravitational covariant ones, could be
restored with the introduction of higher derivative corrections
containing gauge invariant Riemann tensor. These corrections have
coefficients proportional to inverse powers of cosmological constant
so that such theories do not have naive flat limit. But it is
perfectly possible to have a limit when both cosmological term and
gravitational coupling constant simultaneously go to zero in such a
way that only interactions with highest number of derivatives survive.
So the crucial point is the existence of cubic
higher derivative spin $s-s-2$ vertex containing (linearized) Riemann
tensor and two massless spin $s$ particles in flat Minkowski space.
For spin $s=3$ case an appropriate candidate has been
constructed recently in \cite{BL06} (see also \cite{BLS08} where this
vertex was reconsidered and an appropriate one for $s=4$ case has been
constructed). And in our recent paper \cite{Zin08} we have shown that
deformation of this vertex into $(A)dS$ space indeed can reproduce (at
least in linear approximation) standard minimal gravitational
interaction.

Besides gravitational interaction one more classical and important
test for any high spin theory is electromagnetic interaction. The
problem of switching on of such interaction for massless high spin
particles looks very similar to the problem with gravitational
interactions. Namely, if one replace ordinary partial derivatives by
the gauge covariant ones the resulting Lagrangian looses its gauge
invariance and this non-invariance (arising due to non-commutativity
of covariant derivatives) is proportional  to field strength of vector
field. In this, for the massless field with $s \ge 3/2$ in flat
Minkowski space  there is no possibility to restore gauge invariance
by adding non-minimal terms to Lagrangian and/or modifying gauge
transformations. But such restoration becomes possible if one goes to
$(A)dS$ space with non-zero cosmological constant. By the same reason
as in the gravitational case, such theories do not have naive flat
limit, but it is possible to consider a limit where both cosmological
constant and electric charge simultaneously go to zero so that only
highest derivative non-minimal terms survive. Again it is crucial to
have some higher derivative cubic $s-s-1$ vertex containing e/m field
strength. 

In this paper we give an example of such restoration for the case of
massless spin 2 particles. In Section 1, motivated by results of
\cite{BLS08}, we (re)consider cubic third derivative 2-2-1 vertex and
show that interaction Lagrangian indeed could be written in terms of
field strength of vector field (so it is trivially invariant under
vector field gauge transformations) while spin 2 fields gauge
transformations have exactly the same form as in \cite{BLS08}. The
structure of interaction Lagrangian and gauge transformations suggests
transition to so called frame-like formulation where one uses general
non-symmetric second rank tensors for the description of spin 2
particles and in Section 2 we reconstruct the same vertex in such
formalism. At last, in Section 3 we consider deformation of this
vertex into $(A)dS$ space and show that it is indeed possible to
produce minimal e/m interaction at least in linear approximation.

\section{Cubic vertex 2-2-1}

In this Section we (re)consider cubic three derivative 2-2-1 vertex
motivated by results of \cite{BLS08}. In-particular, from the results
of this paper it follows that one needs at least two different spin 2
fields in order such vertex can be constructed. So we consider two
symmetric second rank tensors $h_{\mu\nu}{}^i$, $i=1,2$ and vector
field $A_\mu$. We take sum of the standard kinetic terms for all three
fields as our free Lagrangian:
\begin{equation}
{\cal L}_0 = \frac{1}{2} \partial_\mu h_{\alpha\beta}{}^i \partial_\mu
h_{\alpha\beta}{}^i - (\partial h)_\mu{}^i (\partial h)_\mu{}^i +
(\partial h)_\mu{}^i \partial_\mu h^i - \frac{1}{2} \partial_\mu h^i
\partial_\mu h^i - \frac{1}{4} A_{\mu\nu}{}^2
\end{equation}
This Lagrangian is invariant under usual  gauge transformations:
\begin{equation}
\delta h_{\mu\nu}{}^i = \partial_\mu \xi_\nu{}^i + \partial_\nu 
\xi_\mu{}^i, \qquad \delta A_\mu = \partial_\mu \lambda
\end{equation}

The most general (up to total divergence) cubic vertex with three
derivatives  could be written (schematically):
$$
{\cal L}_1 \sim \partial^3 h_1 h_2 A \oplus \partial^2 h_1 \partial 
h_2 A \oplus \partial h_1 \partial^2 h_2 A \oplus h_1 \partial^3 h_2 A
$$
Correspondingly, we choose the following ansatz for modification of
gauge transformations with vector parameters:
$$
\delta_1 A \sim \partial h_1 \partial \xi_2 \oplus \partial h_2 
\partial \xi_1, \quad \delta_1 h_1 \sim \partial A \partial \xi_2,
\quad \delta_1 h_2 \sim \partial A \partial \xi_1
$$
As for the corrections for $\lambda$-transformations, all of them (up
to possible change of variables) are trivial (i.e. leave sum of
kinetic terms invariant). Indeed, it is easy to check that
transformations
$$
\delta h_{\mu\nu}{}^i = \varepsilon^{ij} [ c_1 R_{\mu\nu}{}^j + c_2
g_{\mu\nu} R^j ] \lambda 
$$
where $R_{\mu\nu}{}^i$ --- linearized Ricci tensor, leave free
Lagrangian invariant and do not correspond to any interaction vertex.
It contradicts with \cite{BLS08}, but it is exactly what one needs for
possible application to electromagnetic interactions, because in this
case the Lagrangian being trivially invariant under 
$\lambda$-transformations can be written in terms of gauge invariant
field strength for the vector field. Then, the requirement that the
Lagrangian be invariant under corrected $\xi_\mu{}^i$-transformations 
leads to the following cubic vertex:
\begin{eqnarray}
M^2 {\cal L}_1 &=& \varepsilon_{ij} A^{\mu\nu} [ \partial_\mu 
h_{\alpha\beta}{}^i \partial_\alpha h_{\beta\nu}{}^j + \frac{1}{2}
\partial_\alpha h_{\beta\mu}{}^i \partial_\alpha h_{\beta\nu}{}^j
- \partial_\alpha h_{\beta\mu}{}^i \partial_\beta h_{\alpha\nu}{}^j
- \frac{1}{2} \partial_\mu h_{\alpha\beta}{}^i \partial_\nu 
h_{\alpha\beta}{}^j + \nonumber \\
 && + \partial_\mu h_{\nu\alpha}{}^i (\partial h)_\alpha{}^j +  
\frac{1}{2} (\partial h)_\mu{}^i (\partial h)_\nu{}^j -
(\partial h)_\mu{}^i \partial_\nu h^j - \partial_\mu h_{\nu\alpha}{}^i
\partial_\alpha h^j + \frac{1}{2} \partial_\mu h^i \partial_\nu h^j ]
\end{eqnarray}
Here, using the fact that cubic bosonic three derivative vertex must
have coupling constant with dimension $1/m^2$, we introduce
appropriate mass scale $M$. Corrections to gauge transformations look
as follows:
\begin{eqnarray}
M^2 \delta h_{\mu\nu}{}^i &=& - \varepsilon^{ij} [ \frac{1}{2} (
A_{\mu\alpha} \partial_{[\alpha} \xi_{\nu]}{}^j + A_{\nu\alpha} 
\partial_{[\alpha} \xi_{\mu]}{}^j ) + \frac{1}{d-2} g_{\mu\nu}
A_{\alpha\beta} \partial_\alpha \xi_\beta{}^j ] \nonumber \\
M^2 \delta A_\mu &=& \varepsilon_{ij} \partial_\alpha h_{\beta\mu}{}^i
\partial_{[\alpha} \xi_{\beta]}{}^j
\end{eqnarray}
Note, that they have exactly the same form as in \cite{BLS08}.

The structure of interaction Lagrangian and gauge transformations 
suggests transition to so called "frame-like" formulation where one
uses general (non-symmetric) second rank tensor for description of
spin 2 particle. So, before turning on to possible electromagnetic
interactions, we reconstruct the same cubic vertex in such
formulation.

\section{Frame-like second order formulation}

As is well known, for symmetric second rank tensor it is impossible to
construct gauge invariant object out of first derivatives of this
field. But if our tensors $h_{\mu\nu}{}^i$ are non-symmetric, we can
choose gauge transformations to be:
$$
\delta h_{\mu\nu}{}^i = \partial_\mu \xi_\nu{}^i
$$
As a result, we can easily construct gauge invariant tensors ("Lorentz
connections"):
\begin{equation}
\Omega_{\mu,\alpha\beta}{}^i = \frac{1}{2} [ \partial_\mu (
h_{\alpha\beta}{}^i - h_{\beta\alpha}{}^i ) - \partial_\alpha (
h_{\mu\beta}{}^i + h_{\beta\mu}{}^i ) + \partial_\beta ( 
h_{\mu\alpha}{}^i + h_{\alpha\mu}{}^i ) ]
\end{equation}
which are antisymmetric on last two indices 
$\Omega_{\mu,\alpha\beta}{}^i = - \Omega_{\mu,\beta\alpha}{}^i$.
But making transition from symmetric tensors to general ones, we
introduce additional degrees of freedom (namely, antisymmetric parts
of our tensors). To compensate this difference, we introduce
additional local transformations:
$$
\delta h_{\mu\nu}{}^i =  \eta_{\mu\nu}{}^i, \qquad
\eta_{\mu\nu}{}^i = - \eta_{\nu\mu}{}^i
$$
In this, $\delta  \Omega_{\mu,\alpha\beta}{}^i = \partial_\mu
\eta_{\alpha\beta}{}^i$. Now the free Lagrangian could be written as:
\begin{equation}
{\cal L}_0 = \frac{1}{2} ( \Omega_{\mu,\alpha\beta}{}^i
\Omega_{\alpha,\mu\beta}{}^i - \Omega_\beta{}^i \Omega_\beta{}^i ) -
\frac{1}{4} A_{\mu\nu} A_{\mu\nu}
\end{equation}
where $\Omega_\beta{}^i = g^{\mu\alpha} \Omega_{\mu,\alpha\beta}{}^i$.

In terms of these variables the interaction Lagrangian takes the form:
\begin{equation}
M^2 {\cal L}_1 = \varepsilon_{ij} A^{\mu\nu} [ \frac{1}{2} 
\Omega_{\mu,\alpha\beta}{}^i \Omega_{\nu,\alpha\beta}{}^i 
- \Omega_{\alpha,\mu\beta}{}^i \Omega_{\beta,\nu\alpha}{}^j
+ \Omega_{\alpha,\mu\nu}{}^i \Omega_\alpha{}^j
+ \Omega_\mu{}^i \Omega_\nu{}^j ]
\end{equation}
It is trivially (by construction) invariant under the 
$\xi_\mu{}^i$-transformations, while invariance under
$\eta_{\mu\nu}{}^i$-transformations requires non-trivial deformations:
\begin{eqnarray}
M^2 \delta_1 h_{\mu\nu}{}^i &=& - \varepsilon^{ij} [ A_{\mu\alpha}
\eta_{\alpha\nu}{}^j + A_{\nu\alpha} \eta_{\alpha\mu}{}^j + 
\frac{1}{d-2} g_{\mu\nu} A^{\alpha\beta} \eta_{\alpha\beta}{}^j ]
\nonumber \\
M^2 \delta_1 A_\mu &=& \varepsilon^{ij} \Omega_{\mu,\alpha\beta}{}^i
\eta_{\alpha\beta}{}^j
\end{eqnarray}
In order to go back to symmetric second rank tensors, one has
supplement $\xi_\mu{}^i$ transformations with $\eta_{\mu\nu}{}^i$-ones
where $\eta_{\mu\nu}{}^i = - \frac{1}{2} ( \partial_\mu \xi_\nu{}^i -
\partial_\nu \xi_\mu{}^i)$ reproducing results of previous Section.

Note also, that if we formally separate indices into "local" and
"world" ones then the Lagrangian could be rewritten in a rather
suggestive form:
\begin{equation}
M^2 {\cal L}_1 = \frac{1}{4} ( \delta_a{}^\mu \delta_b{}^\nu -
\delta_a{}^\nu \delta_b{}^\mu) \varepsilon_{ij} [
4 \Omega_\mu{}^{ac,i} \Omega_\nu{}^{bd,j} A^{cd}
+ \Omega_\mu{}^{cd,i} \Omega_\nu{}^{cd,j} A^{ab}
+ \Omega_\mu{}^{ab.i} \Omega_\nu{}^{cd,j} A^{cd} ]
\end{equation}

\section{Deformation to $(A)dS$}

In this Section we consider transition from flat Minkowski space to
constant curvature $(A)dS_d$ ones and show that the three derivative
cubic vertex considered above is exactly what one needs to compensate
the non-invariance that arises when one introduce minimal
electromagnetic interaction for massless spin 2 field. We will use the
following convention on covariant derivatives:
\begin{equation}
 [ D_\mu, D_\nu ] v_\alpha = R_{\mu\nu,\alpha\beta} v^\beta = - \kappa
( g_{\mu\alpha} v_\nu - g_{\nu\alpha} v_\mu ), \qquad
\kappa = \frac{2 \Lambda}{(d-1)(d-2)}
\end{equation}

Now, due to non-commutativity of covariant derivatives, $\Omega^i$ are
not invariant under gauge transformations with vector parameters:
$$
\delta \Omega_{\mu,\alpha\beta}{}^i = \kappa (g_{\mu\alpha}
\xi_\beta{}^i - g_{\mu\beta} \xi_\alpha{}^i )
$$
Resulting non-invariance of free Lagrangian:
$$
\delta {\cal L}_0 = - \kappa (d-2) \Omega^\alpha{}^i
\xi_\alpha{}^i
$$
could be compensated by adding:
\begin{equation}
\Delta {\cal L}_0 = \frac{\kappa(d-2)}{2} [ h_{\alpha\beta}{}^i
h_{\beta\alpha}{}^i - h^i h^i ]
\end{equation}
In this, total Lagrangian is invariant under transformations with
tensor parameters as well.

Let us introduce minimal e/m interaction by replacing $(A)dS$ 
covariant derivatives by fully covariant ones, e.g.:
\begin{equation}
\nabla_\mu \xi_\alpha{}^i = D_\mu \xi_\alpha{}^i - e_0 
\varepsilon^{ij} A_\mu \xi_\alpha{}^j, \qquad
[ \nabla_\mu, \nabla_\nu ] \xi_\alpha{}^i = [ D_\mu, D_\nu ] 
\xi_\alpha{}^i  - e_0 \varepsilon^{ij} A_{\mu\nu} \xi_\alpha{}^j
\end{equation}
As a result, additional non-invariance under the transformations with
vector parameters appear:
$$
\delta \Omega_{\mu,\alpha\beta}{}^i = - \frac{e_0}{2} \varepsilon^{ij}
[ A_{\mu\alpha} \xi_\beta{}^j - A_{\mu\beta} \xi_\alpha{}^j -
A_{\alpha\beta} \xi_\mu{}^j ]
$$
This leads to non-invariance of free Lagrangian:
$$
\delta_\xi {\cal L}_0 = \frac{e_0}{2} \varepsilon_{ij} A^{\mu\nu} [
\Omega_{\alpha,\mu\nu}{}^i \xi_\alpha{}^j + 2 \Omega_\mu{}^i
\xi_\nu{}^j ]
$$
At the same time, invariance of free Lagrangian under the
transformations with tensor parameters is also broken:
$$
\delta_\eta {\cal L}_0 =  \frac{e_0}{2} \varepsilon_{ij} A^{\mu\nu} [
2 h_{\alpha\mu}{}^i \eta_{\nu\alpha}{}^j + h^i \eta_{\mu\nu}{}^j ]
$$

Now we add the third derivative cubic vertex, obtained in the previous
Section, where all derivatives in the Lagrangian and gauge
transformations are replaced by the covariant ones. Non-invariance of
this vertex due to non-commutativity of covariant derivatives has the
form (up to the terms quadratic in $A_{\mu\nu}$):
\begin{equation}
M^2 \delta_\xi {\cal L}_1 = - \kappa (d-3) \varepsilon_{ij} A^{\mu\nu}
[ \Omega_{\alpha,\mu\nu}{}^i \xi_\alpha{}^j + 2 \Omega_\mu{}^i
\xi_\nu{}^j ] - 2 \kappa \varepsilon_{ij} A^{\mu\nu}
\Omega_{\mu,\nu\alpha}{}^i \xi_\alpha{}^j
\end{equation}
For $e_0 = 2\kappa(d-3)/M^2$ first term compensate non-invariance of
free Lagrangian, while to compensate the second term we add finally:
\begin{equation}
\Delta {\cal L}_1 = a_0  \varepsilon_{ij} A^{\mu\nu} h_{\mu\alpha}{}^i
h_{\nu\alpha}{}^j, \qquad
\delta A_\mu = b_0 \varepsilon_{ij} h_{\mu\alpha}{}^i \xi_\alpha{}^j 
\end{equation}
When for $a_0 = 2 \kappa$, $b_0 = 2 \kappa$ all variations cancel.
Moreover, resulting Lagrangian, as we have explicitly checked, turns
out to be invariant under transformations with tensor parameters as
well.

\section*{Conclusion}

Thus we have shown that for massless spin 2 particles in $(A)dS$ space
it is indeed possible to switch on minimal electromagnetic
interactions supplemented by third derivative non-minimal
interactions. But similar possibility exists for the massive particles
in flat Minkowski space \cite{FPT92,KZ97,DW01d,PR08}. By the same
reasoning it is natural to suggest that there must exist a limit where
both mass and electric charge go to zero so that some non-minimal
higher derivative terms survive. Non-minimal terms similar to the ones
considered here appeared in \cite{FPT92}, while in our investigation
\cite{KZ97}, based on gauge invariant description of massive spin 2
particles, it turned out that to restore gauge invariance after
switching on minimal e/m interactions it is enough to introduce 
corrections with two derivatives only. So it appears that the relation
between massless particles in $(A)dS$ space and massive ones in flat
Minkowski space is not so simple and straightforward as it seemed. Any
way, this question certainly deserves further study.


\begin{thebibliography}{10}

\bibitem{AD79}
C.~Aragone, S.~Deser
{\it "Consistency problem of hypergravity",}
Phys. Lett. {\bf B86} (1979) 161.

\bibitem{WF80}
D.~de~Wit, D.~Z. Freedman
{\it "Systematics of higher-spin gauge fields",}
Phys. Rev. {\bf D21} (1980) 358.

\bibitem{BBD85}
F.~A. Berends, G.~J.~H. Bugrers, H.~van Dam
{\it "On the theoretical problems in constructing interactions
involving higher-spin massless particles",}
Nucl. Phys. {\bf B260} (1985) 295.

\bibitem{Por08}
M.~Porrati
{\it "Universal Limits on Massless High-Spin Particles",}
arXiv:0804.4672.

\bibitem{FV87}
E.~S. Fradkin, M.~A. Vasiliev
{\it "On the gravitational interaction of massless higher-spin
fields",}
Phys. Lett. {\bf B189} (1987) 89.

\bibitem{FV87a}
E.~S. Fradkin, M.~A. Vasiliev
{\it "Cubic interaction in extended theories of massless higher-spin
fields",}
Nucl. Phys. {\bf B291} (1987) 141.

\bibitem{BL06}
N.~Boulanger, S.~Leclercq
{\it "Consistent couplings between spin-2 and spin-3 massless
fields",}
JHEP {\bf 0611} (2006) 034, arXiv:hep-th/0609221.

\bibitem{BLS08}
N.~Boulanger, S.~Leclercq, P.~Sundell
{\it "On The Uniqueness of Minimal Coupling in Higher-Spin Gauge
Theory",} arXiv:0805.2764.

\bibitem{Zin08}
Yu.~M. Zinoviev
{\it "On spin 3 interacting with gravity",} arXiv:0805.2226.

\bibitem{FPT92}
S.~Ferrara, M.~Porrati, V.~L. Telegdi
{\it "$g=2$ as the natural value of the tree-level gyromagnetic ratio
of elementary particles",}
Phys. Rev. {\bf D46} (1992) 3529.

\bibitem{KZ97}
S.~M. Klishevich, Yu.~M. Zinoviev
{\it "On electromagnetic interaction of massive spin-2 particle",}
Phys. Atom. Nucl. {\bf 61} (1998) 1527, arXiv:hep-th/9708150.

\bibitem{DW01d}
S.~Deser, A.~Waldron
{\it "Inconsistencies of massive charged gravitating higher spins",}
Nucl. Phys. {\bf B631} (2002) 369, arXiv:hep-th/0112182.

\bibitem{PR08}
M.~Porrati, R.~Rahman
{\it "Intrinsic Cutoff and Acausality for Massive Spin 2 Fields
Coupled to Electromagnetism",} arXiv:0801.2581.

\end{thebibliography}
\end{document}